\documentclass[aps,pre,
reprint,
 notitlepage,
 showpacs,floatfix,nofootinbib,showkeys,
 superscriptaddress
 ]{revtex4-2}

 \usepackage{subfigure}
 \usepackage{amssymb}
 \usepackage{amsfonts}
 \usepackage{amsmath}
 \usepackage{amsthm}
 \usepackage{epsfig}
 \usepackage{float}
 \usepackage[usenames,dvipsnames]{color}
 \usepackage[latin1]{inputenc}
 \usepackage{enumitem}
 
\usepackage{url} 
 
 \usepackage{graphics,graphicx,xcolor,subfigure}

 \usepackage[hidelinks,unicode=true]{hyperref}
 \hypersetup{colorlinks=true,
 	linkcolor=blue,
 	urlcolor=blue,
 	citecolor=blue,
 	pdfhighlight=/N
 }
 
 \usepackage{comment}
 \usepackage[normalem]{ulem}
 \usepackage{microtype}

\newcommand{\av}[1]{\langle {#1} \rangle}

\begin{document}

\title{Consequences of non-Markovian healing processes on epidemic models with recurrent infections on networks}

\author{José Carlos M. Silva}\thanks{These authors contributed equally to this work.}
\affiliation{Departamento de F\'{\i}sica, Universidade Federal de Vi\c{c}osa, 36570-900 Vi\c{c}osa, Minas Gerais, Brazil}

\author{Diogo H. Silva}\thanks{These authors contributed equally to this work.}
\affiliation{Instituto de Ci\^{e}ncias Matem\'{a}ticas e de Computa\c{c}\~{a}o, Universidade de S\~{a}o Paulo, S\~{a}o Carlos, SP 13566-590, Brazil}

\author{Francisco A. Rodrigues}
\affiliation{Instituto de Ci\^{e}ncias Matem\'{a}ticas e de Computa\c{c}\~{a}o, Universidade de S\~{a}o Paulo, S\~{a}o Carlos, SP 13566-590, Brazil}

\author{Silvio C. Ferreira}
\affiliation{Departamento de F\'{\i}sica, Universidade Federal de Vi\c{c}osa, 36570-900 Vi\c{c}osa, Minas Gerais, Brazil}
\affiliation{National Institute of Science and Technology for Complex Systems, 22290-180, Rio de Janeiro, Brazil}
 
\begin{abstract} 
Infections diseases are marked by recovering time distributions which can be far from the exponential one associated with Markovian/Poisson processes, broadly applied in epidemic compartmental models. In the present work, we tackled this problem by investigating a susceptible-infected-recovered-susceptible model on networks with $\eta$ independent infectious compartments (SI$_{\eta}$RS), each one with a Markovian dynamics, that leads to a Gamma-distributed recovering times. We analytically develop a theory for the epidemic lifespan on star graphs with a center and $K$ leaves, which mimic hubs on networks, showing that the epidemic lifespan scales with a non-universal power-law $\tau_{K}\sim K^{\alpha/\mu\eta}$ plus logarithm corrections, where $\alpha^{-1}$ and $\mu^{-1}$ are the mean waning immunity and recovering times, respectively. Compared with standard SIRS dynamics with $\eta=1$ and the same mean recovering time, the epidemic lifespan on star graphs is severely reduced as the number of stages increases. In particular,  the case $\eta\rightarrow\infty$ leads to a finite lifespan. Numerical simulations support the approximated analytical calculations. For the SIS dynamics, where no immunity is conferred ($\alpha\rightarrow\infty$), numerical simulations show that the lifespan increases exponentially with the number of leaves, with a nonuniversal rate that decays with the number of infectious compartments. We investigated the SI$_{\eta}$RS dynamics on power-law networks with degree distribution $P(K)\sim k^{-\gamma}$.  When $\gamma<5/2$ and the epidemic processes are ruled by a maximum $k$-core activation, the alteration of the hub activity time does not alter either the epidemic threshold or the localization pattern. For $\gamma>3$, where hub mutual activation is at work, the localization is reduced but not sufficiently to alter the threshold scaling with the network size. Therefore, the activation mechanisms remain the same as in the case of Markovian healing.

\end{abstract}

\maketitle

\section{Introduction}

The compartmental epidemic models have been modified to address different spreading phenomena including migratory mobility in metapopulation models~\cite{Colizza2007a, Barrat2007}, 
different geographical scales and locations~\cite{Guilherme2020}, dynamics of vector-borne diseases~\cite{Angel2017}, role of  multiple strains in cooperative interactions~\cite{Poletto2015}, epidemic events occurring at different time scale~\cite{Ventura2021}, and behavioral response to epidemic scenarios~\cite{Steinegger2020,Mozhgan2022,SILVA2023}. The basic stochastic methods to deal with compartmental epidemic models consider memoryless processes using Markovian dynamics~\cite{Kiss2017} where the ongoing events are independent of the history~\cite{Cox1977}. In this approach transitions between compartments follow Poisson processes with exponential distributions of interevent times that allow simplified mathematical and computational approaches.  However, real epidemic data processes are consistent with non-Markovian, thus trajectory-dependent,  processes~\cite{Kiss2017,Keeling2011,Romualdo2015}. For a contemporary example, the COVID-19 viral loads are not consistent with exponentially distributed infectious times~\cite{Jones2021}, and its period of incubation, which corresponds to the time elapsed between infection and onset of symptoms, was estimated as a random Weibull variable~\cite{JING2020}.

Non-Markovian epidemic models are characterized by non-exponential distributions for interevent times~\cite{Keeling2011,Kiss2017}, and epidemic outcomes, such as the epidemic threshold, basic reproduction number, and epidemic incidence, may be significantly altered compared to the Markovian case~\cite{Cator2013, Van_Mieghen2013, Van_Mieghem2019, Wilkinson2018, Feng2019}. While the importance of non-Markovian dynamics is well established for susceptible-infected-recovered (SIR) models~\cite{Keeling2011, Rohani2005, Kiss2017}, which do not involve reinfections, the role of non-Markovianity in models with recurrent infections remains challenging~\cite{Van_Mieghem2019, Van_Mieghen2013, Michele2017, Feng2019}. For example, in susceptible-infected-susceptible (SIS) dynamics, which can have an active steady state, considering Gamma or Weibull distributions for infection times and Poisson processes (exponentially distributed times) for recovery, the mean-field prediction for the epidemic threshold is modified~\cite{Van_Mieghen2013, Van_Mieghem2019}. However, non-Markovian distributions for recovery times do not alter the mean-field epidemic threshold of SIS dynamics~\cite{Cator2013}. In this mean-field case, since the recovery time is exponentially-distributed, an effective infection rate can be defined, and an equivalence between Markovian and non-Markovian dynamics can be established~\cite{Michele2017}.

In non-recurrent dynamics, such as susceptible-infected-recovered (SIR) and susceptible-exposed-infected-recovered (SEIR) models, for a fixed average recovery period, decreasing the variance of the recovery time distribution increases the probability that the disease will spread, and this probability is maximized with a nonrandom recovery period~\cite{Wilkinson2018}, a feature observed in models with multiple recovery stages~\cite{Rohani2005}. In a simple version of these models, an infected individual passes through $\eta$ independent and identically exponentially distributed stages before recovering, which leads to a Gamma distribution for the infectious period~\cite{Keeling2011}. The Gamma-distributed recovery and latent periods produce a rich bifurcation structure, which could be associated with the effects of vaccination and demographic trends in the seasonality of whooping cough in England and Wales~\cite{Rohani2008}. The number of stages in such epidemic events is a key parameter~\cite{Rohani2008} in this broad diversity of scenarios.

The mean-field results for SIS dynamics, which suggest similarity between non-Markovian and Markovian recovery processes given that the infection process is Markovian, have led to reduced interest in models with multiple stages for recurrent infection at a constant rate. However, the activation mechanisms of SIS and SIRS dynamics on highly heterogeneous networks can be much more intricate than indicated by mean-field approaches~\cite{Kitsak2010, Castellano2012, Goltsev2012, Boguna2013, Mata2015, Sander2016, Silva2019, Jose_Carlos2022}. A central point is the feedback mechanism of activation between hubs and their nearest neighbors~\cite{Boguna2013, Sander2016}, which can lead to different types of localized activation~\cite{Kitsak2010, Castellano2012, Sander2016}, where the actual epidemic dynamics can be triggered by a long-range activation mechanism, depending on the network structure and the nature of the epidemic model~\cite{Durret2009, Boguna2013, Sander2016}. In particular, the inclusion of waning immunity in the susceptible-infected-recovered-susceptible (SIRS) model markedly alters the scenario for networks with a power-law degree distribution $P(k) \sim k^{-\gamma}$ with exponent $\gamma > 3$, where SIS dynamics presents an asymptotically null epidemic threshold~\cite{Durret2009, Boguna2013, Ferreira2012}, while the SIRS model exhibits a finite threshold~\cite{Sander2016, Jose_Carlos2022}.

In the present work, we investigated the SIRS epidemic model with $\eta$ statistically identical infectious stages, hereafter called SI$_\eta$RS, on star graphs composed of a center and $K$ leaves, mimicking isolated hubs in networks. Using a modified version of discrete-time dynamics for SIRS on star graphs proposed by Ferreira et al.~\cite{Sander2016} (see also Bogu\~n\'a et al.~\cite{Boguna2013} for the theory of the SIS model), we derive an approximate analytical expression for the epidemic lifespan, considering an average healing time of $1/\mu$ and a small infection rate $\lambda \ll \mu$. The analytical expression yields an epidemic lifetime $\langle\tau_{K}\rangle \sim K^{-\frac{\alpha}{\eta \mu}}$ for finite $\alpha$, along with logarithmic corrections. In particular, for finite $\alpha$ and $\eta \rightarrow \infty$, this lifetime is finite. Stochastic simulations in which a hub is modeled as an isolated star graph support the analytical result for finite $\alpha$, but the theory does not capture the dependence with $\eta$ in the limit of instantaneous waning immunity, which corresponds to the SIS model. We also investigated the transition from an epidemic-free to an endemic state of the SI$_\eta$RS model on different networks. Despite the remarkable differences in the epidemic lifespan of isolated hubs as the number of infectious states increases, stochastic simulations on power-law networks with sizes up to $10^7$ nodes lead only to a reduction in epidemic localization and a displacement of the epidemic threshold for $\gamma > 3$, with no significant difference observed for scale-free networks with $\gamma < 3$. For the case of random regular (RR) networks with $P(k) = \delta_{k,6}$, a reduction in the epidemic threshold is reported. Therefore, the alterations in the epidemic lifespan on isolated hubs for non-Markovian dynamics are not sufficient to change the activation mechanisms of the Markovian SIRS dynamics on networks.

The remainder of this paper is organized as follows. In Sec.~\ref{sec:model}, the model is described, highlighting some consequences of the $\eta$ stages on the recovery time distribution. In Sec.~\ref{sec:theory}, the average lifetime feedback mechanism and the hub's mutual infection time are analytically determined, grounded in the theoretical framework of Refs.~\cite{Durret2009, Boguna2013, Sander2016}, and compared with (statistically exact) stochastic simulations. The critical SI$_\eta$RS dynamics is investigated on power-law networks using stochastic simulations in Sec.~\ref{sec:results}, while our conclusions and prospects are summarized in Sec.~\ref{sec:conclusion}. Three appendices with simulation and analytical technical details complement the article.

\section{The SI$_\eta$RS Model}
\label{sec:model}

\begin{figure}[htb]
	\includegraphics[width=0.9\linewidth]{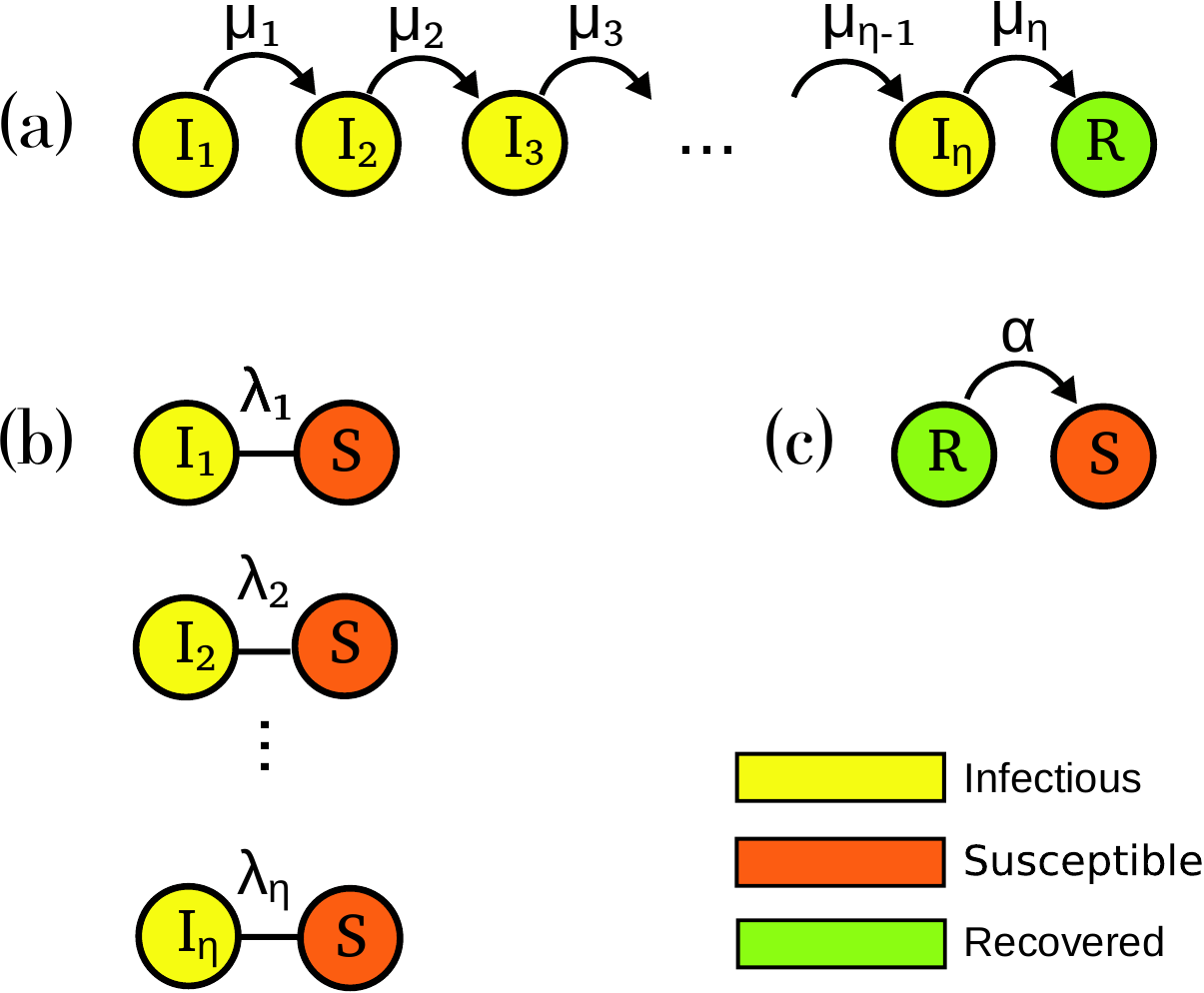}
	\caption{Schematic representation of (a) infectious state evolution until the recovering state,  (b) transmission processes of the disease, and (c) waning of immunity returning to the susceptible state in SI$_\eta$RS epidemic model with $\eta$ compartments.}
	\label{fig:model}
\end{figure}

We define a general SI$_\eta$RS model in which the recovery process sequentially passes through $j=1, 2, \ldots, \eta$ infectious stages, represented by I$_1$, I$_2$, $\ldots$, I$_\eta$, as schematically shown in Fig.~\ref{fig:model}(a). The transition rates between infectious states $j$ and $j+1$ (for $j < \eta$) are time-independent and given by $\mu_j$, while the transition from $j=\eta$ to the recovered state R occurs at rate $\mu_\eta$. The transitions between subsequent infectious stages and recovery processes occur spontaneously. An infectious individual in stage $j=1,\ldots,\eta$ transmits the disease to a susceptible contact at constant rates $\lambda_1, \ldots, \lambda_\eta$, respectively, as depicted in Fig.~\ref{fig:model}(b). A newly infected node always starts in stage $\eta=1$. A recovered individual returns to the susceptible state at rate $\alpha$, as shown in Fig.~\ref{fig:model}(c). For simplicity, we will consider the same infection and recovery rates for all stages: $\lambda_{j} = \lambda$ and $\mu_j = \mu \eta$. This choice is suitable for comparing different values of $\eta$, as the average recovery time is given by $\langle \tau \rangle = 1/\mu$, irrespective of $\eta$. The SI$_\eta$RS dynamics includes the particular cases of SIR and SIS models with $\eta$ infectious stages when $\alpha \rightarrow 0$ and $\alpha \rightarrow \infty$, respectively. 

The convolution of identical and independent exponentially distributed transition times between subsequent infectious states  given by $\psi(t_j)=\eta \mu \exp(-\eta \mu t_j)$ leads to a Gamma distribution for the total recovering time,
\begin{equation}
\tau=\sum_{1}^{\eta} t_j,
\end{equation}
given by~\cite{Durret1996,Keeling2011}
\begin{equation}
	\psi_\text{rec}(\tau) = \frac{(\eta\mu)^\eta}{(\eta-1)!} \tau^{\eta-1}e^{-\eta\mu \tau}.
	\label{eq:Gamma}
\end{equation}

\begin{figure}[htb]
	\includegraphics[width=0.95\linewidth]{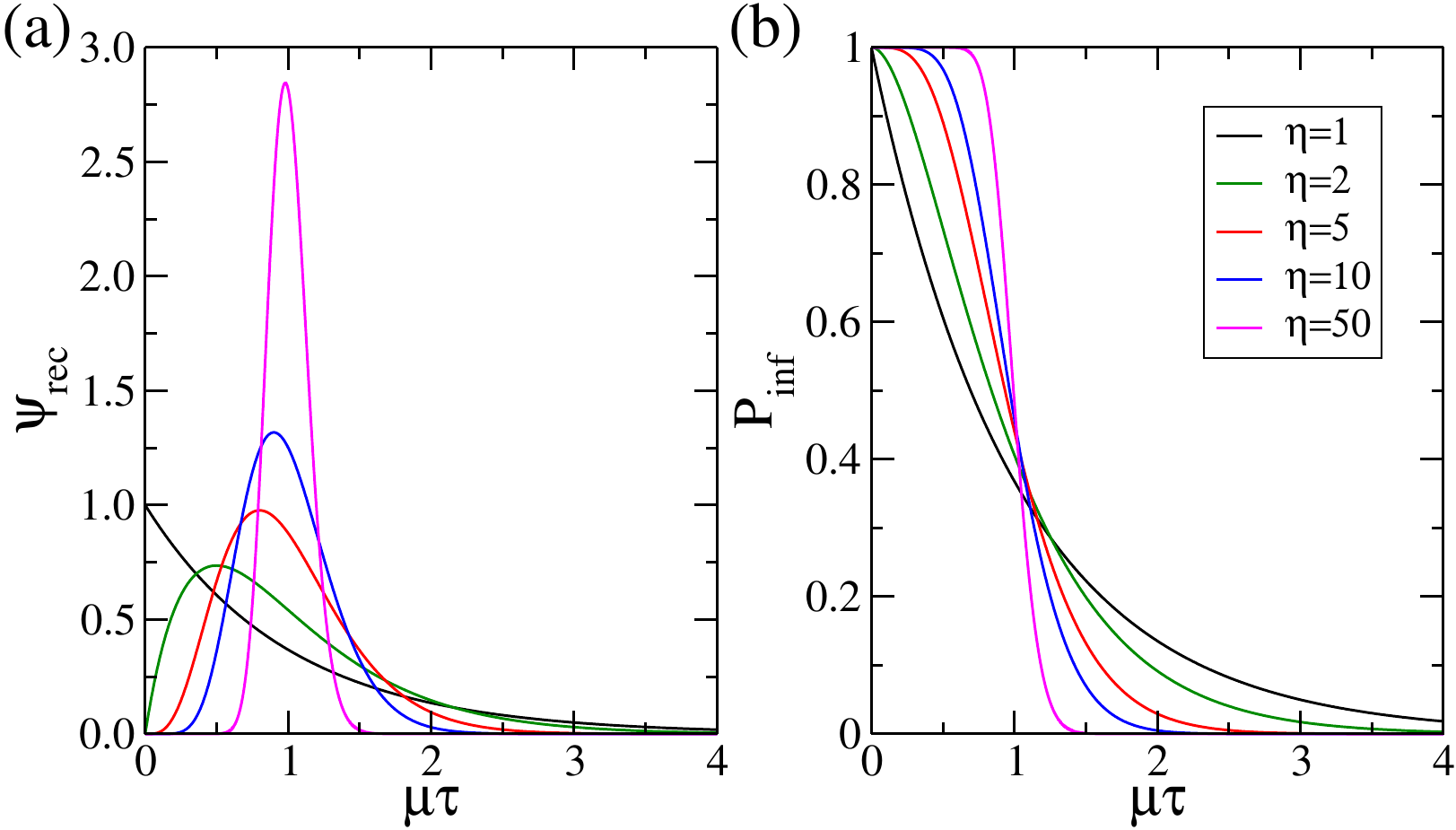}
	\caption{(a) Distribution of recovering period ($\psi_{\text{rec}}$) and (b) probability of remaining infected ($P_{\text{inf}}$) until a time $\tau$ for healing processes with $\eta$ stages.}
	\label{fig:infection_distribution}
\end{figure}

As the number of stages increases, the distribution approaches a Dirac delta function centered at \(\tau = \langle \tau \rangle = 1/\mu\), as shown in Fig.~\ref{fig:infection_distribution}(a) for \(\eta = 1, 2, 5, 10,\) and \(50\). The probability of being infected up to time \(\tau\) since infection is given by
\begin{equation}
	P_\text{inf}(\tau) = 1 - \int_{0}^{\tau} \psi_\text{rec}(\tau') d\tau',
	\label{eq:Pinf}
\end{equation}  
which converges to a Heaviside step function as \(\eta \to \infty\)~\cite{Rohani2005}; see Fig.~\ref{fig:infection_distribution}(b). Household data for whooping cough fit better with a Gamma-distributed recovery period than with an exponential one~\cite{Rohani2008}.

\section{SI$_\eta$RS dynamics on star graphs}
\label{sec:theory}

The transition from a disease-free state to an endemic phase can be triggered by different activation mechanisms depending on the network structure, which is related to the degree exponent in random networks~\cite{Castellano2012,Boguna2013,Sander2016}. A key aspect of understanding the activation mechanism for $ \gamma > 3 $ is determining the epidemic lifespan on star graphs, which model sparse hubs in a network~\cite{Durret2009,Boguna2013,Sander2016}. This mechanism relies on feedback between the center (hub) and its leaves (neighbors). The hub infects several leaves, which can then reinfect the hub, even at low infection rates per edge. If the infection duration is long enough for an event starting at the hub to activate other hubs far from the source, a long-range mutual infection mechanism is at play.

\subsection{Epidemic lifespan}
\label{subsec:lifepsan}

To estimate the epidemic lifetime of a star graph for SI$_\eta$RS models, we adapt the theory of Ref.~\cite{Sander2016} considering discrete-time dynamics where each time step has the following sequence. 
\begin{itemize}[leftmargin=16pt]
	\item[i.)] At $t=0$, the center is infected in the stage $j=1$ and the $K$ leaves are susceptible.
	
	\item[ii)] At $t=t_1=1/\mu$, the center is recovered and $n_1$ leaves are infected (stage $j=1$) according to a binomial distribution
	\begin{equation}
		P_1(n_1|K) = \binom{K}{n} p_1^{n_1}(1-p_1)^{K-n_1},
		\label{eq:P_1}
	\end{equation}
	where $p_1$ is the probability that the center infects a susceptible neighbor before healing and is computed as the complementary probability that the individual passes through all $\eta$ compartments without healing, which is given by
	\begin{equation}
		p_1=1-\prod_{j=1}^{\eta} \left[\frac{\mu_j}{\lambda_j+\mu_j}\right],
		\label{eq:pmu_j}
	\end{equation}
	since transitions $j\rightarrow j+1$ and infections are independent Poisson processes with rates $\mu_j$ and $\lambda_j$, respectively.  For the case of interest  $\mu_j=\eta\mu$ and $\lambda_j=\lambda$, Eq.~\eqref{eq:pmu_j} becomes     	
	\begin{equation}
		p_1=1-\left[\frac{\mu \eta}{\lambda+\mu\eta}\right]^\eta.
		\label{eq:pmu}
	\end{equation}
	The limit cases for Eq.~\eqref{eq:pmu} of $\eta=1$ (Markovian healing time) and $\eta=\infty$ (deterministic healing time) are $p_1=\lambda/(\lambda+\mu)$ and $p_1=1-\exp(-\lambda/\mu)$, respectively. In both cases, $p_1\approx \lambda/\mu$ in the limit of interest $\lambda\ll\mu$.
	
	\item[iii)] At $t=t_{1}+t_{2}$, in which $t_{2}$ follows an exponential distribution $\rho_{2}=\alpha\exp\left(-\alpha t_{2}\right)$,  the center returns to susceptible state while $0\le n_2\le n_1$ leaves remain infected with probability
	\begin{equation}
		P_2(n_2|n_1) = \binom{n_1}{n_2} p_2^{n_2} (1-p_2)^{n_1-n_2},
		\label{eq:P_2}
	\end{equation}
	where $p_2=P_\text{inf}(t_2)$, given by Eqs.~\eqref{eq:Pinf}  and \eqref{eq:Gamma}.
	
	\item[iv)] At the time $t=t_{1}+t_{2}+t_{3}$, where $t_{3}=1/\mu$, all leaves become susceptible synchronously, and the center is reinfected, returning to the initial configuration, with probability
	\begin{equation}
		q(n_2) = 1-(1-p_1)^{n_2},
	\end{equation}
	that is the probability that at least one leaf tried to infect the center. Otherwise, with probability $1-q(n_2)$ the dynamic ends.
\end{itemize}

So, the probability that the epidemic process remains active at $t=t_3=2/\mu+t_2$ is given by
\begin{align}
	Q_K(t_2) & = \sum_{n_1=1}^{K}\sum_{n_2=1}^{n} P_1(n_1|K)P_2(n_2|n_1)q(n_2) \nonumber \\
	         & = 1-[1-p_1^2p_2]^K,
\label{eq:Q_K}	
\end{align}
where we used Eqs.~\eqref{eq:P_1} and \eqref{eq:P_2} and performed the algebraic manipulations using Newton binomial formula. Averaging over $t_2$, one has the probability that the epidemic survives the sequence of  steps i) to iv) is given by
\begin{equation}
	\bar{Q}_{K}=1-\alpha\int_{0}^{\infty}e^{-\alpha t_{2}}(1-p_1^2 p_2)^{K}dt_{2}.
	\label{eq:QKmed}
\end{equation}

\begin{figure}[hbt]	
\includegraphics[width=0.85\linewidth]{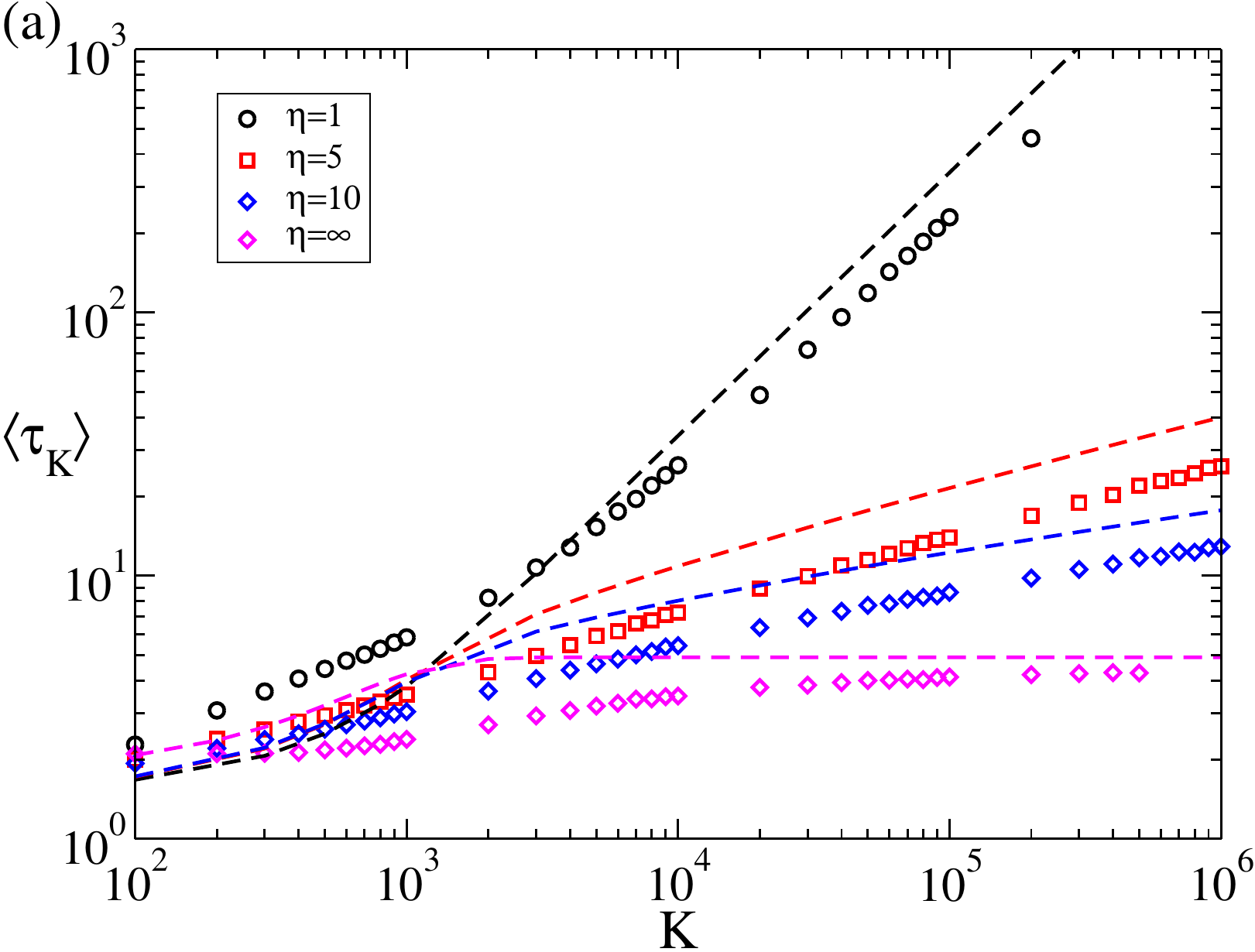}\\
\includegraphics[width=0.85\linewidth]{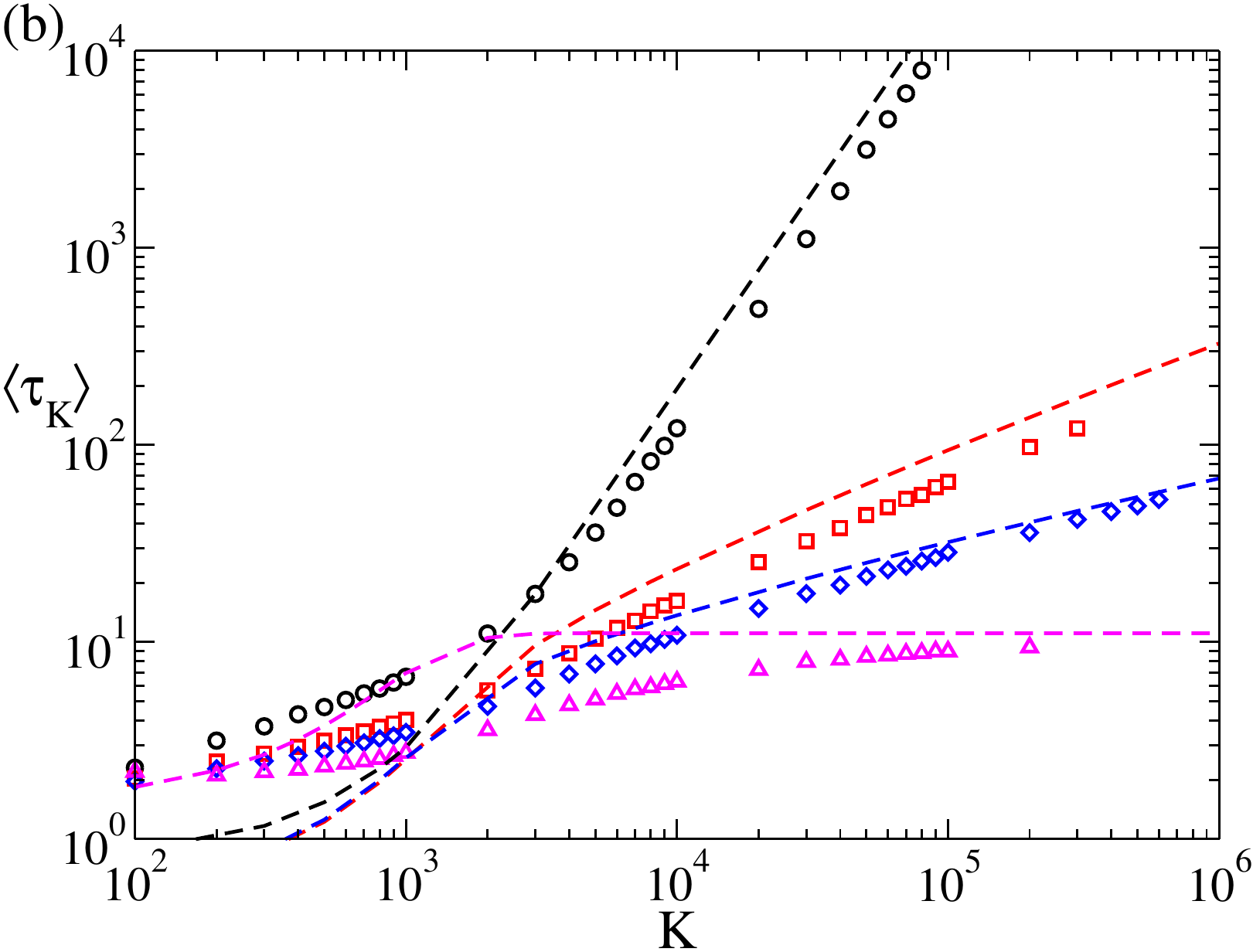}\\
\includegraphics[width=0.85\linewidth]{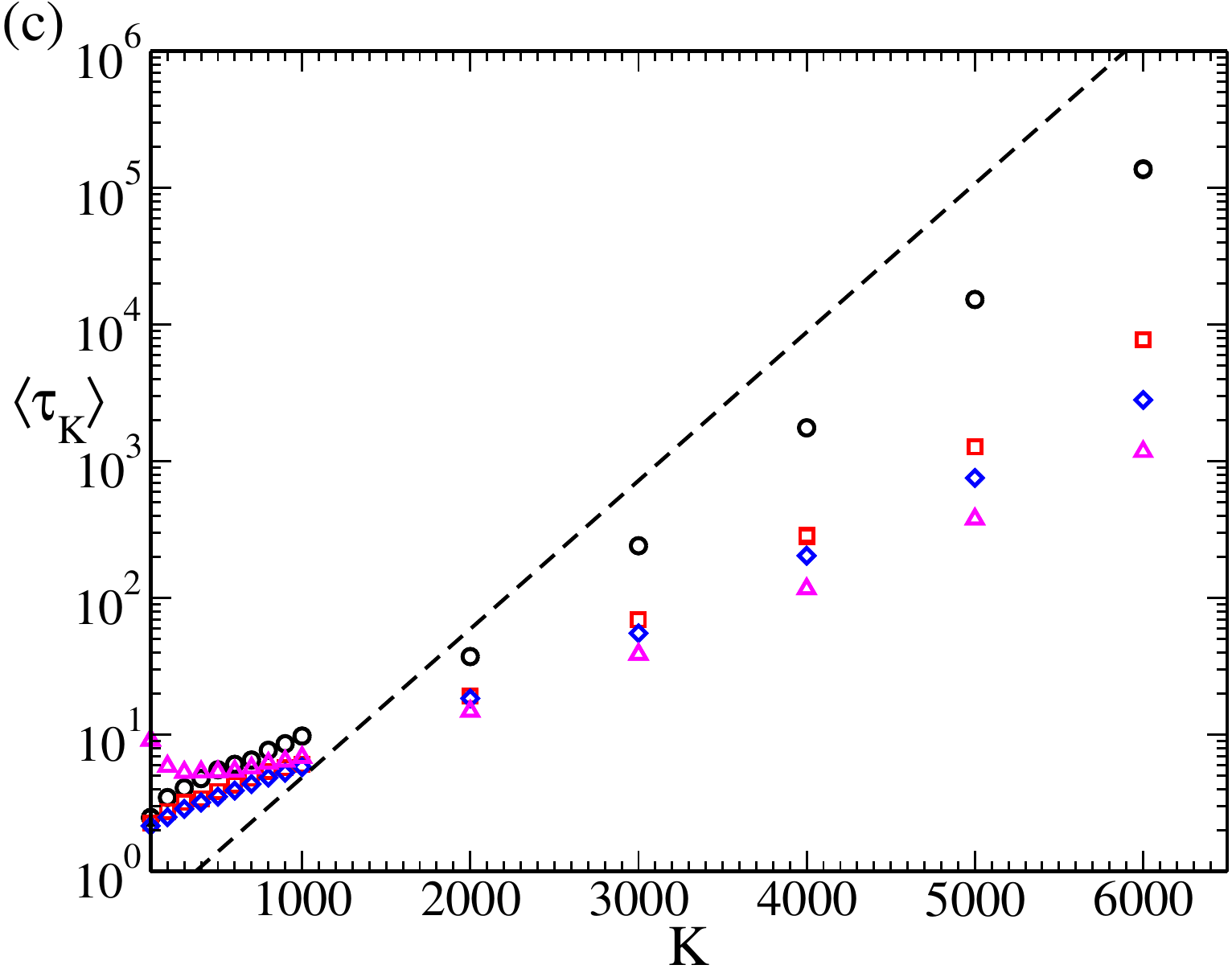}
\caption{Epidemic lifespan $\av{\tau_{K}}$ as function of the star graph size $K$, for different number of infectious stages $\eta=1$, $\eta=5$, $\eta=10$, and $\eta =\infty$ (deterministic recovering time). The parameters of the dynamics are $\lambda=0.05$, $\mu=1$, (a) $\alpha=1$, (b) $\alpha=2$ and (c) $\alpha= \infty$ (SIS). Symbols are simulations while dashed lines correspond to Eqs. \eqref{eq:tau_Kmed} and \eqref{eq:tauK_eta_infty}   for  finite and infinite $\eta$, respectively.}
\label{fig:lifespan_star}
\end{figure}

The number of steps $n_s$ that the dynamics remain active follows a geometric distribution given by 
\begin{equation}
	P_s(n_s) = (\bar{Q}_K)^{n_s-1}(1-\bar{Q}_K),
\end{equation} 
that means the dynamics is reactivated $n_s-1$ times in a row and ends at the step $n_s$.  So the activity lifetime of a hub and its $K$ nearest neighbors, $\av{\tau_{K}}$  assumes the form  
\begin{equation}
	\av{\tau_{K}}=\tau(\av{n_s}) = \frac{\tau}{1-\bar{Q}_{K}}.
	\label{eq:tau_Kmed}
\end{equation}
in which, $\tau=t_{1}+\av{t_{2}}+t_{3}= 2/\mu+1/\alpha$ is the average duration of a cycle.

To determine the epidemic lifespan $\av{\tau_{K}}$ is necessary to evaluate the integral given by Eq.~\eqref{eq:QKmed}. For a finite number of stages $\eta$, the asymptotic limit can be computed by applying the saddle-point method (see appendix~\ref{app:saddle}), which results in    
\begin{eqnarray}
	\av{\tau_{K}} \sim  (Kp_1^{2})^{\frac{\alpha}{\eta\mu}}\left( \ln \left(Kp_1^{2}\right)\right)^{\frac{\eta-1}{\eta}\frac{\alpha}{\mu}},
	\label{eq:tauK_asymp}
\end{eqnarray}
in the limit of large $K$. In particular, when $\eta=1$, Eq.~\eqref{eq:tauK_asymp} corresponds to the result of Ref.~\cite{Sander2016} for the Markovian SIRS dynamics, where $\av{\tau_K}\sim K^{\alpha/\mu}$.

In the limit case $\eta\rightarrow\infty$, the saddle-point method can not be applied to evaluate Eq.~\eqref{eq:QKmed}. In this limit, the healing processes becomes deterministic $\psi_\text{rec}(\tau)=\delta(\tau-1/\mu)$ and $p_2=P(t_2)$ becomes a step function
\begin{eqnarray}
	p_{2}=\left\{\begin{array}{ll}
		1,~\text{ if }~ t_{2}<1/\mu\\
		0,~\text{ if }~ t_{2}>1/\mu.
	\end{array}\right.	
	\label{eq:p_2step}
\end{eqnarray}
Plugging Eq.~\eqref{eq:p_2step} into  \eqref{eq:QKmed}, one obtains
\begin{equation}
\begin{aligned}
1-\bar{Q}_K
 = & \alpha\int_{0}^{1/{\mu}}e^{-\alpha t_{2}} e^{-p_1^2 K} dt_{2} +\alpha\int_{1/\mu}^{\infty}e^{-\alpha t_{2}}dt_{2}  \\
 = & e^{-p_1^2 K}\left(1-e^{-\alpha/\mu}\right) + e^{-\alpha/\mu},
\end{aligned}
	\label{eq:1menosQK}
\end{equation}
where we have used $(1-p_1^2)^K\approx\exp(-Kp_1^2)$ valid for $p_1\ll 1$. Now, substituting Eq.~\eqref{eq:1menosQK} in Eq.~\eqref{eq:tau_Kmed} we obtain
\begin{equation}
	\av{\tau_{K}}=\frac{\tau}{e^{-Kp_1^{2}}\left(1-e^{-\alpha/\mu}\right)+e^{-\alpha/\mu}},
	\label{eq:tauK_eta_infty}
\end{equation}
which assumes the constant value $\av{\tau_K} =\tau \exp\left(\alpha/\mu\right)$ for $K\rightarrow \infty$ and $\alpha/\mu$ finite.

The theoretical predictions obtained by the numerical integration of Eq.~\eqref{eq:QKmed}  or analytical expression Eq.~\eqref{eq:tauK_eta_infty} are in very good agreement with continuous-time stochastic simulations (see appendix~\ref{app:stochastic_simulations}) on a star graph for large $K$ and distinct values of $\eta$, as shown in Fig.~\ref{fig:lifespan_star}(a) and (b). The only difference is the prefactor which is not a relevant issue given the approximated discrete-time dynamics used for derivation of the analytical results.

The limit case  $\alpha\rightarrow \infty$, corresponding to the SIS dynamics, can be directly obtained  using $\lim_{\alpha\rightarrow \infty} \alpha\exp(-\alpha t) =\delta(t)$ in Eq.~\eqref{eq:QKmed}:
\begin{equation}
	\begin{aligned}
		1-Q_K  & = \int_{0}^{\infty}\delta(t_{2}) (1-p_1^2p_2) dt_{2} \\
		& = (1-p_1^2)^K\simeq e^{-K\frac{\lambda^2}{\mu^2}},
	\end{aligned}
	\label{eq:1menosQKSIS}
\end{equation} 
where the last approximation is valid $\lambda/\mu\ll 1$ where $p_1\approx \lambda/\mu$. So, in the SIS dynamics, the analytical average lifetime is independent of $\eta$ and given by
\[\av{\tau_{K}} \approx \frac{2}{\mu}e^{K\frac{\lambda^2}{\mu^2}},\]
in agreement with the Markovian SIS result of Ref.~\cite{Boguna2013}. Simulations agree with the exponential increase in the star graph size. However, the exponential's slope varies slightly with $\eta$ as can be seen in Fig. \ref{fig:lifespan_star}(c). Thus, the feedback activation mechanism of Markovian SIS dynamics on star graphs~\cite{Boguna2013} is not significantly altered by the Gamma distribution of the recovering period and one expects that the activation mechanism by hubs in power-law networks is not significantly altered. Indeed, this result is in agreement with  Ref.~\cite{Van_Mieghen2013}, in which, at a mean-field level, the epidemic threshold is independent of the recovery time distribution once the healing times are exponentially distributed.

 \subsection{Mutual infection time} 
\label{subsec:inf_mutual}

\begin{figure}[!htb]
	\includegraphics[scale=0.45]{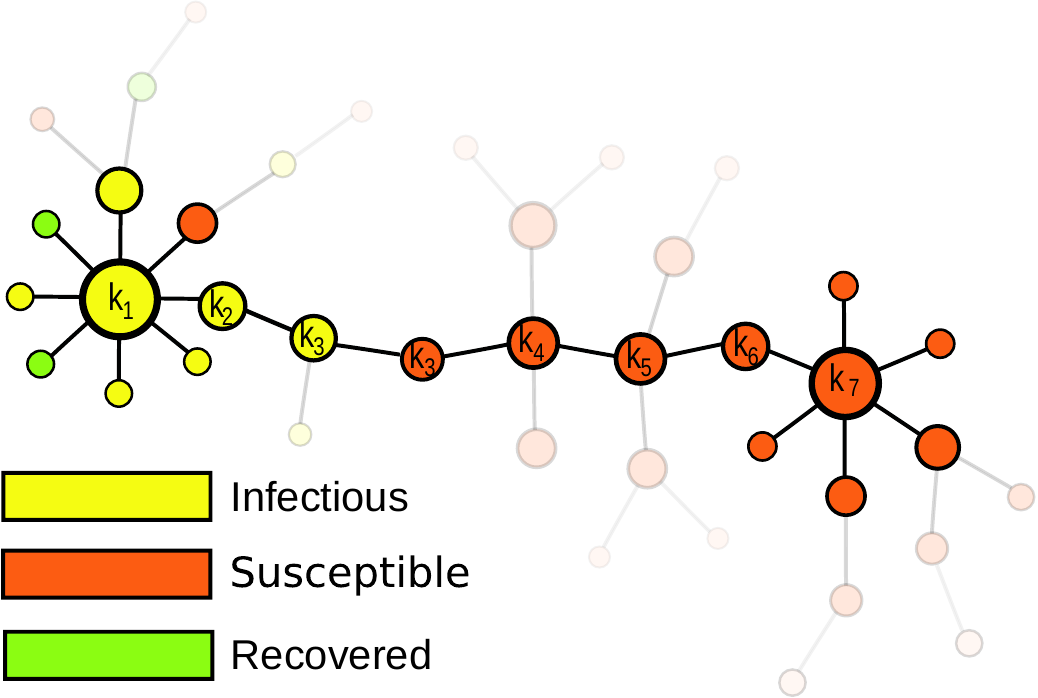}
	\caption{An illustrative case of communication between hubs.
		The infection starts at the center of the left hub of degree $K_{1}$ (source) and travels through this path to reach the right hub of degree $K_{7}$ (target). The shaded nodes represent other connections that are not relevant for the infection propagation. } 			
	\label{fig:PATH}
\end{figure}

Another essential aspect of the activation mechanism is the long-range mutual infection of hubs, which is driven by fluctuations~\cite{Durret2009,Boguna2013,Sander2016}. To estimate the average time $\langle \tau_{KK'} \rangle$ for infections in a hub of degree $K$ to reach another hub of degree $K'$, we analyze a path of length $d$ connecting a source of degree $K$ and a target of degree $K'$ (see figure~\ref{fig:PATH}). In this calculation, the source does not recover, representing an upper bound for the actual probability of long-range infection. The probability that a node in the path infects its neighbor before recovering is given by Eq.~\eqref{eq:pmu}. Thus, an infection that starts at the source reaches the target with probability $p_1^{d-1}$, which corresponds to an effective infection rate given by $\lambda p_1^{d-1}$.  

For random uncorrelated networks with $N$ nodes, the average distance between nodes of degree $K$ and $K'$ is given by~\cite{Holyst}
\begin{equation}
	d=1+\frac{\ln (N \av k/KK')}{\ln \kappa},
	\label{eq:eq10}
\end{equation} 
in which,
\begin{equation}
	\kappa= \frac{\av{k^{2}}-\av{k}}{\av{k}}.
	\label{eq:kappa}
\end{equation}
Therefore, the estimated average time required for the source to activate the target is given by,  
\begin{equation}
	\av{\tau_{KK'}}=\frac{1}{\lambda p_1^{d-1}}= \frac{1}{\lambda} \left(\frac{N\av{k}}{KK'}\right)^{b(\lambda, \eta)},
	\label{eq:tauKKprime}
\end{equation}
in which,  
\begin{equation}
	b(\lambda,\eta)=-\frac{ \ln p_1 }{\ln \kappa}\ge 0.
	\label{eq:b_lb_eta}
\end{equation}
Since, in a network, there are other paths linking hubs, $\av{\tau_{KK'}}$ is an upper bound to mutual infection time. This expression is essentially the same derived for Markovian SIS and SIRS dynamics in Refs.~\cite{Boguna2013,Sander2016} with the only difference of using the modified $p_1$. Since in the limit of low infection $\lambda/\mu\ll 1$, $p_1=\lambda/\mu$ independently of $\eta$, we have that the result is indeed equivalent to Refs.~\cite{Boguna2013,Sander2016}.

\section{SI$_\eta$RS dynamics on networks.}
\label{sec:results}

\begin{figure*}[hbt]
	\includegraphics[width=0.7\linewidth]{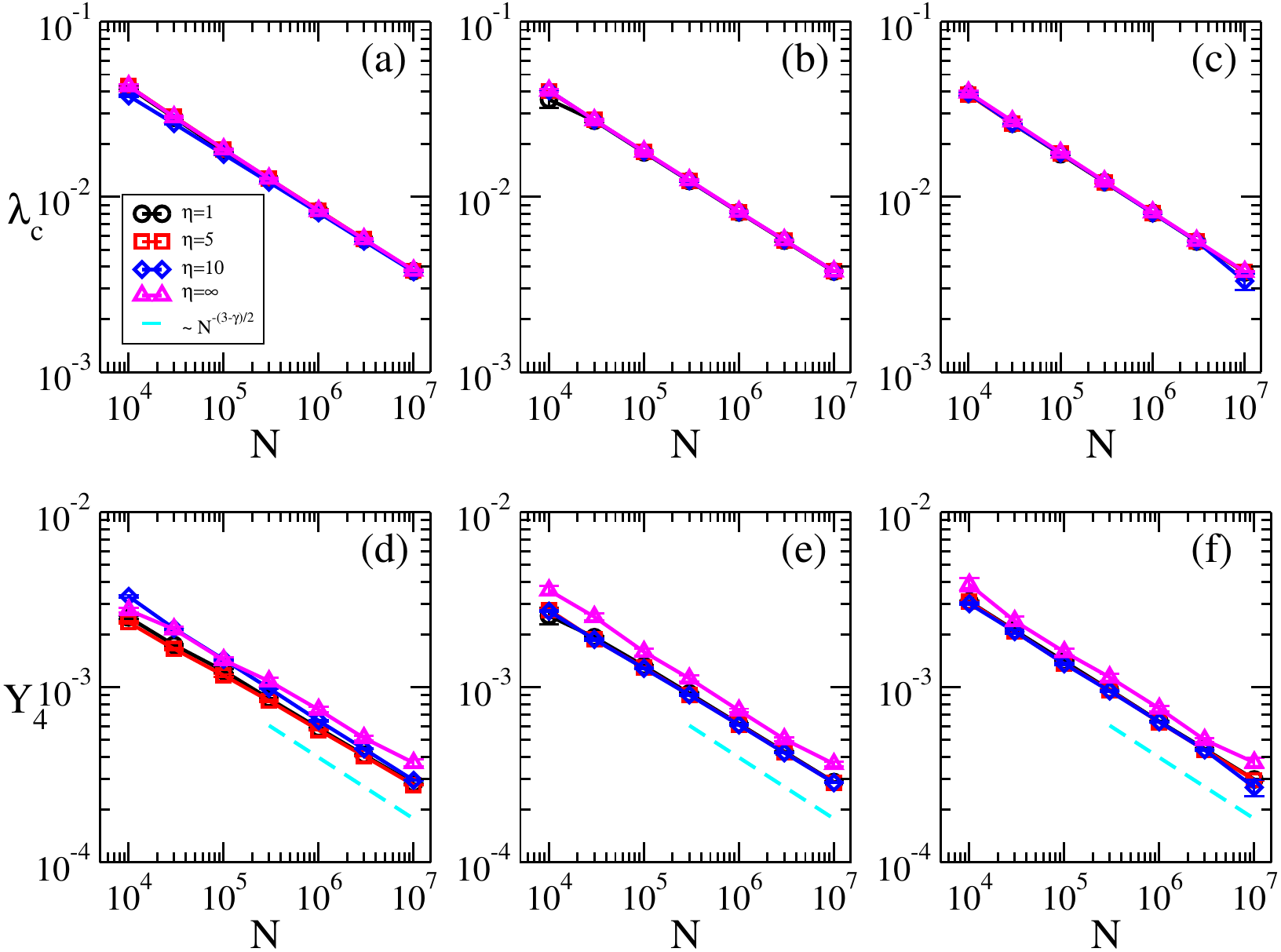}
	\caption{Epidemic threshold (top) and IPR of activity (bottom) as function of network size for (a,d) $\alpha=0.5$, (b,e) $\alpha=1$, and (c,f) $\alpha=2$ and different numbers of stages $\eta$. The dashed lines are power-law decays $Y_{4}\sim N^{(3-\gamma)/2}$ corresponding to a maximum $k$-core localization. The simulations were run on power-law networks with $\gamma=2.3$, $k_{\text{min}}=3$ and  $k_{\text{c}}=2\sqrt{N}$.}
	\label{fig:gamma230}	
\end{figure*}

The analytical results for the epidemic lifetime and mutual infection time indicate that alterations in relation to Markovian spreading may occur in the dynamics since the epidemic lifetime on hubs is modified by the number of infectious stages $\eta$, while the mutual long-range infection time is not. To analyze these possible effects, we consider uncorrelated networks with a power-law degree distribution, $P(k) \sim k^{-\gamma}$, generated with the uncorrelated configuration model (UCM)~\cite{Catanzaro2005}, where an upper cutoff $k_\text{c} \sim \sqrt{N}$ is used to ensure the absence of degree correlations in simple graphs without self-loops or multiple connections. The most connected node of the network will have an average degree $\langle k_{\text{max}} \rangle \sim N^{1/2}$ for $\gamma < 3$ and $\langle k_{\text{max}} \rangle \sim N^{1/(\gamma - 1)}$ for $\gamma \ge 3$~\cite{Boguna2004}. Finally, the scaling of $\kappa$, Eq.~\eqref{eq:kappa}, is given by  
\begin{equation}
	\kappa \sim \left\lbrace\begin{array}{lll}
		\langle k_{\text{max}} \rangle^{(3-\gamma)/2} \sim N^{(3-\gamma)/2}  &, & \gamma < 3\\
		\text{const} &, & \gamma \ge 3.
	\end{array}\right.
	\label{eq:kappa_scaling}
\end{equation}
Plugging Eq.~\eqref{eq:kappa_scaling} into Eq.~\eqref{eq:b_lb_eta}, we obtain $b(\lambda, \eta) \rightarrow 0$, and $\langle \tau_{KK'} \rangle$ increases slower than algebraically for $\gamma < 3$ in the limit of large network size, regardless of $\eta$. Thus, Eq.~\eqref{eq:tauK_asymp}, which holds for finite $\eta$, implies that $\langle \tau_{K} \rangle > \langle \tau_{KK'} \rangle$. This means that an epidemic spreading starting in a hub would remain active long enough to communicate the disease to other hubs, thereby triggering an outbreak.

To explicitly check the effect of the multiple infection stages, we ran stochastic simulations (see appendix~\ref{app:stochastic_simulations}) of SI$_\eta$RS dynamics on UCM networks with $\gamma = 2.3$. For this exponent, the Markovian dynamics is activated by a densely connected core formed of hubs, as identified by an innermost component of $k$-core decomposition, the maximum $k$-core~\cite{Dorogovtsev2006,Castellano2012}. We analyzed the dynamical susceptibility~\cite{Ferreira2012}, defined as $\chi = N\left(\langle \rho^{2} \rangle - \langle \rho \rangle^{2}\right)/\langle \rho \rangle$, as a function of the infection rate. For systems undergoing a transition to absorbing states in complex networks, the position of the principal peak in these curves is an estimate of the epidemic threshold~\cite{Ferreira2012,Mata2015}.
 
When $\gamma = 2.3$, the introduction of infectious stages does not yield any appreciable effect on the threshold for the investigated network sizes (up to $10^{7}$) and values of $\alpha$ adopted, as shown in the top panels of Fig.~\ref{fig:gamma230}. Thus, the same activation mechanism triggered in the maximum $k$-core, as in the SIS model~\cite{Castellano2012,Kitsak2010}, seems to be at work irrespective of $\eta$. To validate this hypothesis, we evaluated a normalized activity vector $\phi_i$ introduced in Ref.~\cite{Diogo2021}, which is given in terms of the probability $\rho_i$ that node $i$ is infected in the quasistationary state (see Appendix~\ref{app:stochastic_simulations}): $\phi_i \propto \rho_i$ and $\sum_i \phi_i^2 = 1$. We compute the inverse partition ratio (IPR)~\cite{Goltsev2012}, defined as $Y_{4} = \sum_i \phi_i^4$. Regardless of $\eta$, we observed $Y_{4} \sim N^{-\left(3 - \gamma\right)/2}$ for all $\alpha/\mu$ considered, as shown in the bottom of Fig.~\ref{fig:gamma230}. This scaling is associated with the maximum $k$-core localization~\cite{Satorras2016}. Note that a finite lifetime of hubs in the case $\eta = \infty$ is not able to alter the scenario.

\begin{figure*}[hbt]
	\includegraphics[width=0.7\linewidth]{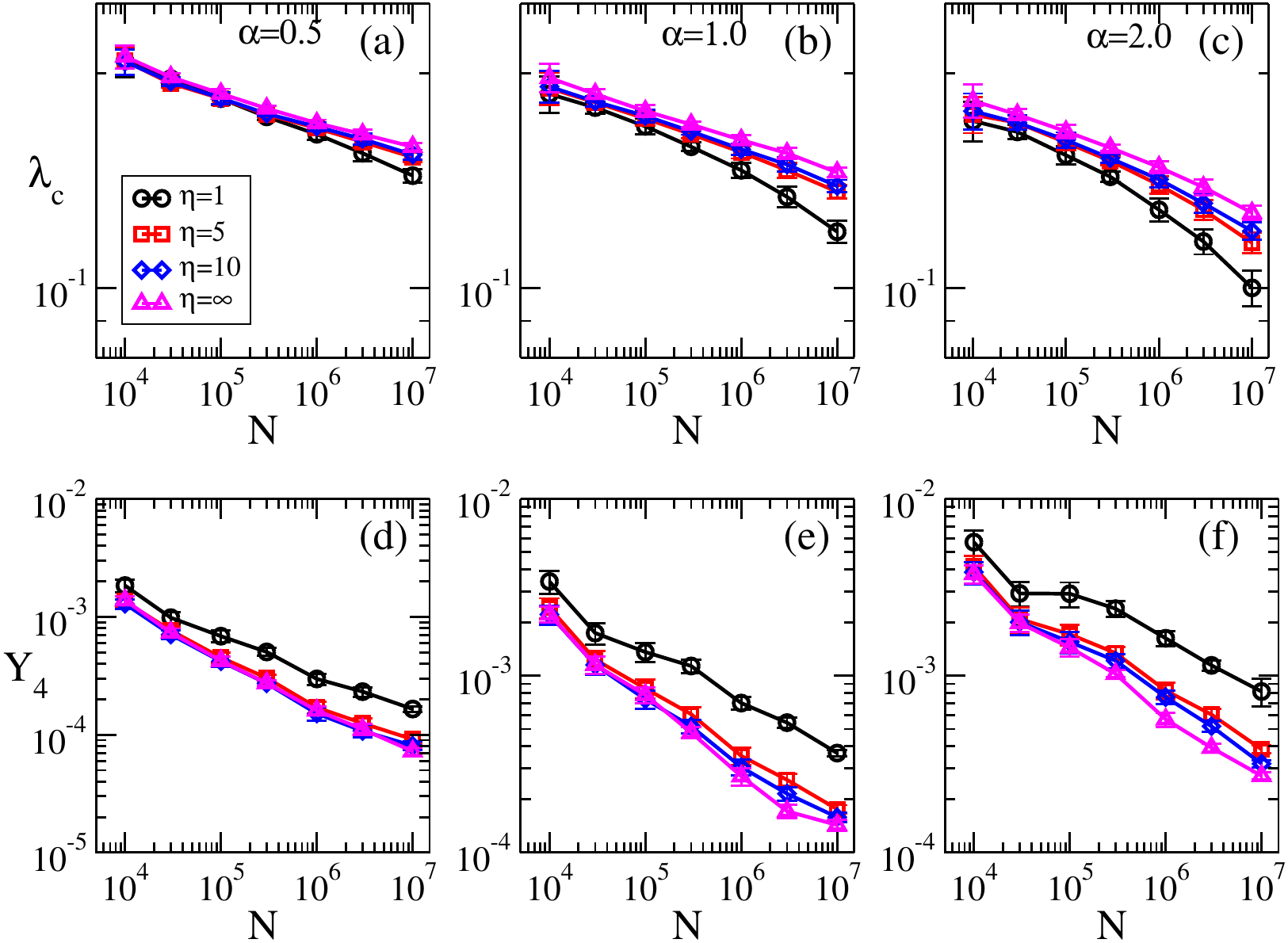}
	\caption{Epidemic thresholds (top) and IPR of activity (bottom) as functions of network size for (a,d) $\alpha=0.5$ , (b,e) $\alpha=1$, and (c,f) $\alpha=2$ for different number of infectious stages $\eta$. The simulations were run on power-law networks with $\gamma=3.5$, $k_{\text{min}}=3$ and natural upper cutoff.}
	\label{fig:gamma350}
\end{figure*}  

For $\gamma > 3$, a long-range mutual activation of hubs may be present in place of the maximum $k$-core activation, as in the case of Markovian SIS~\cite{Boguna2013}. However, this mechanism does not rule the epidemic activation of the Markovian SIRS dynamics, which instead happens collectively, involving an extensive part of the network~\cite{Sander2016}. Since the epidemic lifespan on star graphs decreases with $\eta$, the hubs have shorter-lived activity for $\eta > 1$ compared to the Markovian case $\eta = 1$, while the scaling with network size of mutual infection time is independent of $\eta$. Therefore, the collective activation where $\av{\tau_{KK'}} \gg \av{\tau_{K}}$ remains valid. Figure~\ref{fig:gamma350} presents the analysis of the SI$_\eta$RS dynamics for $\gamma = 3.5$. For $\alpha \le 0.5$, the epidemic threshold is only slightly altered with respect to the Markovian SIRS, as shown in Fig.~\ref{fig:gamma350}(a). The impact of non-Markovian dynamics on the epidemic threshold becomes more evident as $\alpha$ increases, as illustrated in Figs.~\ref{fig:gamma350}(b) and (c), where we observe an increase in the epidemic threshold with $\eta$. This is consistent with the analysis of epidemic lifetime on star graphs since a shorter activity duration on hubs implies slower epidemic spreading. In particular, when $\eta \rightarrow \infty$, from Eq.~\eqref{eq:tauK_eta_infty} and Eq.~\eqref{eq:tauKKprime}, we find that $\av{\tau_{KK'}} \gg 1$, while $\av{\tau_{K}}$ remains finite for $\lambda/\mu \ll 1$. Figure~\ref{fig:gamma350}, however, does not indicate the limit of threshold saturation as a function of size. Indeed, as in the Markovian case~\cite{Sander2016}, this regime is observable, for the investigated values of $\alpha$, only for exceedingly large sizes, much larger than those that can currently be simulated. The saturation can be seen for smaller values of $\alpha$, but in this regime, the effects of multiple stages are very small, as in the case of $\gamma = 2.3$ show i Fig.~\ref{fig:gamma230}.

\begin{figure}[hbt]
	\includegraphics[scale=0.45]{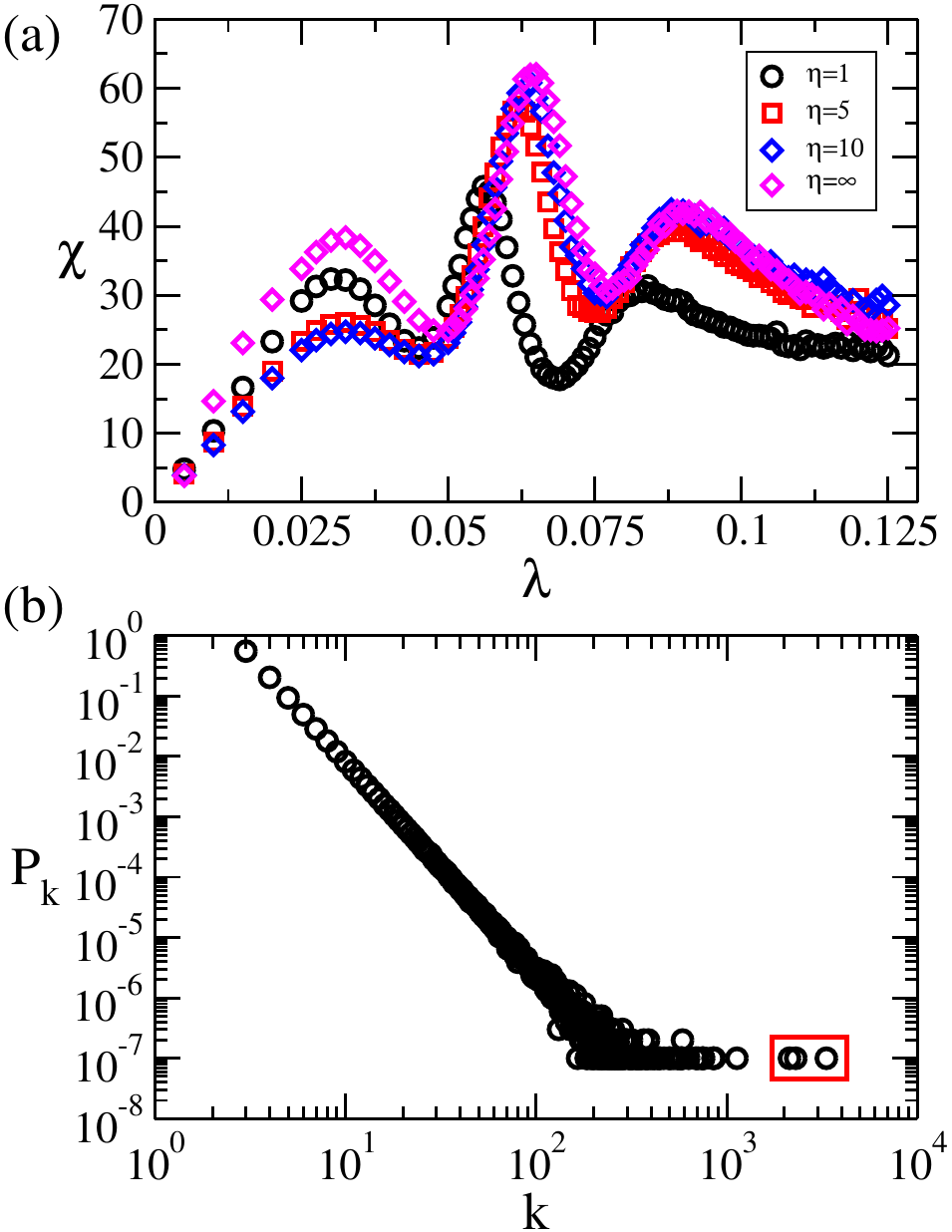}	
	\caption{(a) Susceptibility as a function of the infection rate for different number of infectious stages. Simulations were run in a network with degree distribution shown in (b). The red square highlights the outliers of this network. The other parameters are $N=10^{7}$, $\gamma=3.5$, $k_{\text{min}}=3$, and natural upper cutoff.} 
	\label{fig:gamma350_SIS}
\end{figure}

In the SIS model $(\alpha \rightarrow \infty)$, the existence of hubs can trigger multiple peaks in the susceptibility curves, indicating the activation of different parts of the networks~\cite{Mata2015,Ferreira2012}. Figure~\ref{fig:gamma350_SIS} confirms the existence of multiple peaks in the susceptibility curves for all values of $\eta$ studied in networks with $\gamma > 3$ and the presence of outliers in the degree distribution, as shown in Fig.~\ref{fig:gamma350_SIS}(b). The observable effect is the slight shift of the curves and peaks compared to the Markovian case $\eta = 1$, implying that the localization pattern holds regardless of $\eta$. This result aligns with the effects of the number of stages on the epidemic lifetime shown in Sec.~\ref{subsec:lifepsan}, supporting the resilience of the long-range hub mutual activation mechanisms.

We conclude the results section by considering random regular networks (RRNs), where each node has the same degree $k$ and edges among them are formed randomly without self-loops or multiple connections. A homogeneous mean-field theory to compute the epidemic threshold of the S$I_\eta$RS on RRNs is developed in Appendix~\ref{app:RRN}. Multiple stages do not alter the epidemic threshold in a mean-field approach, given by $\lambda_{\text{c}}/\mu = 1/k$, irrespective of the number of stages $\eta$. This result is consistent with the non-Markovian SIS model~\cite{ROST2020,Cator2013}. However, in stochastic simulations, the introduction of several infectious stages shifts the epidemic threshold to lower values. Incorporating dynamical correlations into the mean-field approach can enhance the precision of the theoretical framework, as seen in the Markovian case~\cite{Jose_Carlos2022}.

\begin{figure*}[hbt]
	\includegraphics[width=0.7\linewidth]{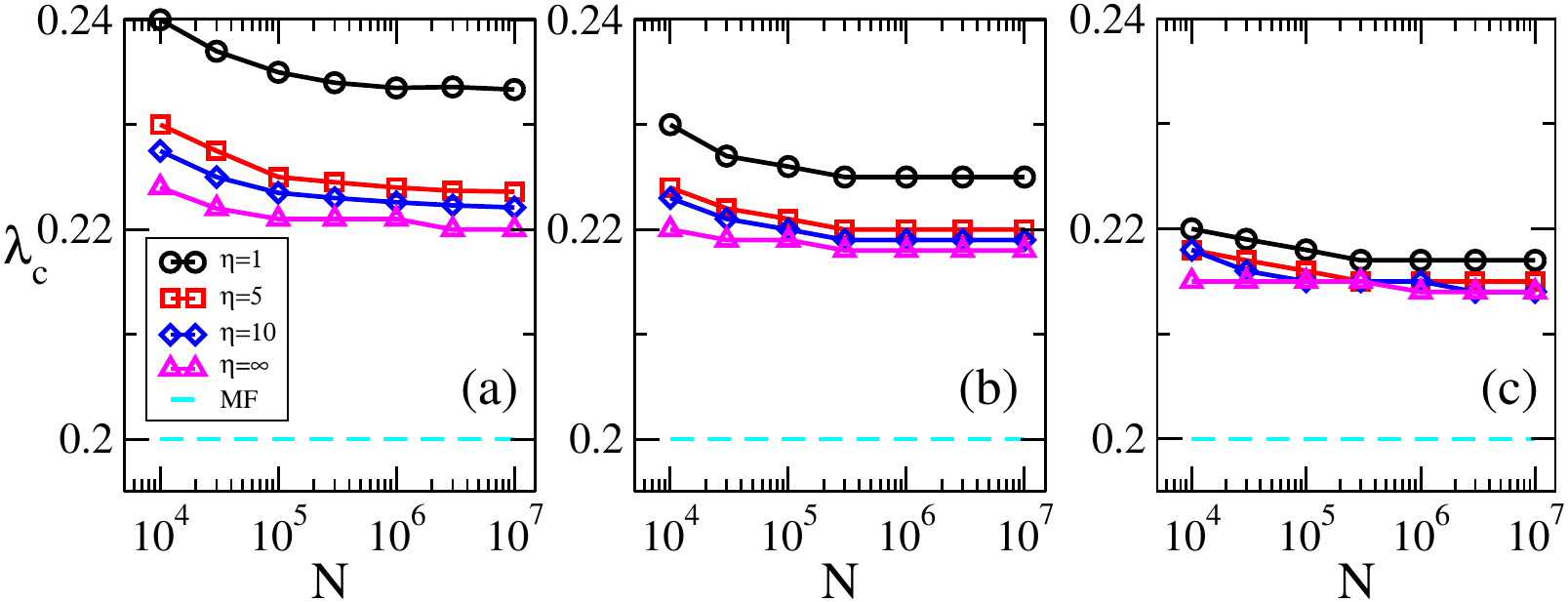}
	\caption{Epidemic threshold as a  function of the network size for a  RR networks with $k=6$ and different number of infectious states $\eta$. The cyan dashed line corresponds to mean-field prediction. We waning immunity rates are (a) $\alpha=0.5$, (b) $\alpha=1$, and (c) $\alpha=2$.}
	\label{fig:RRN}
\end{figure*}

\section{Conclusion}
\label{sec:conclusion}

We investigated the role of non-Markovicity in the recovering processes of the SIRS epidemic model, considering $j=1,\ldots,\eta$ infectious stages that evolve sequentially from $j$ to $j+1$ towards the recovered state $R$. We consider the simplest case where all infectious stages have the same infection rates as well as the transition rates between subsequent stages, implying that the recovery time follows a Gamma distribution~\cite{Keeling2011} with the same average time as the Markovian SIRS model, which has exponentially distributed recovery times. Our focus is on the activation mechanisms that trigger the endemic phase. Therefore, we investigated the epidemic dynamics on star graphs of size $K+1$, which mimic the sparsely distributed hubs in the network and serve as a proxy for determining whether the activation mechanism is governed by the mutual activation of sparsely distributed hubs or by a densely connected set of hubs determined by the maximum $k$-core~\cite{Dorogovtsev2006} of the network.

An approximated analytical calculation based on discrete-time dynamics predicts that the non-Markovicity of the recovering processes drastically alters the SIRS dynamics on star graphs, strongly reducing the average epidemic lifespan compared to the Markovian case. Indeed, the larger the number of infectious stages $\eta$, the lower the scaling of the epidemic lifespan with the star graph size, while the mutual long-range infection among hubs remains essentially unaltered. In particular, the limit $\eta\rightarrow\infty$, which corresponds to a deterministic recovery time, leads to a finite epidemic lifespan even in the limit of infinite star graph size. The analytical results are supported by stochastic simulations for different values of the waning immunity rate $\alpha$. However, for the SIS limit where $\alpha\rightarrow\infty$, the lifespan increases exponentially with the graph size, similar to the Markovian case.

In order the evaluate of effects of the activity lifespan of hubs on the epidemic spreading on networks, we considered stochastic simulations on large random networks with power-law degree distributions in two regimes: $\gamma=2.3$ which is characterized by a maximum $k$-core activation~\cite{Castellano2012,Sander2016,Satorras2016} and $\gamma=3.5$ where $k$-core structure is not present~\cite{Dorogovtsev2006}. In the former, the epidemic localization pattern and, consequently, the epidemic threshold are unaltered by multiple infectious stages in consonance with mean-field analysis for non-Markovian SIS on networks~\cite{Cator2013}. In the latter, a shift in the epidemic threshold towards values smaller than the Markovian ones and a reduction of the epidemic localization are observed, but not sufficient to change the qualitative conclusions reported previously for the Markovian SIRS~\cite{Sander2016,Jose_Carlos2022}, at least for the range of networks size attainable in this work.

In a nutshell, although the non-Markovicity through multiple infectious stages alter significantly the lifespan of localized epidemic activity, it is not relevant to the nature of the activation processes ruling recurrent infection dynamics on the whole networks. As prospects of the present work, we can consider the effects of viral-load~\cite{Van_Mieghem2019} and multiple stages in the recovering compartment~\cite{Jensen2019}.

\appendix

\section{Stochastic simulations}
\label{app:stochastic_simulations}


Since all infectious stages evolve according to Poisson processes, one can perform stochastic simulations of the SI$_\eta$RS epidemic model following similar steps to the optimized Gillespie algorithm presented in Ref.~\cite{Cota2017}. In the investigated model, the rates $\mu_i=\mu\eta$ and $\lambda_j=\lambda$ are uniform across all stages. Thus, we can group all infection events and transitions between infectious stages into broader events as follows. The total number of infected individuals $N_{\text{inf}}$ (regardless of the infectious stage), the total number of edges emanating from them $N_{\text{SI}}$, and the total number of recovered individuals $N_{\text{rec}}$ are computed to determine the probabilities of each group event.

In each time step, one of three broader events is chosen as follows. With probability 
\begin{equation}
	P_{\text{I}} = \frac{\mu \eta N_{\text{inf}}}{\mu\eta N_{\text{inf}} + \lambda N_{\text{SI}} + \alpha N_{\text{rec}}},
	\label{eq:GA_PI}
\end{equation}
one infected node is selected at random. If it is in the last infectious state $j=\eta$, it recovers: I$_\eta \rightarrow$ R. Otherwise, if $j < \eta$, it progresses to the next infectious stage: I$_j \rightarrow$ I$_{j+1}$. Note that recovery and progression through infectious stages are treated equally in terms of rates. 

With probability 
\begin{equation}
	P_{\text{R}} = \frac{\alpha N_{\text{rec}}}{\mu\eta N_{\text{inf}} + \lambda N_{\text{SI}} + \alpha N_{\text{rec}}},
	\label{eq:GA_PR}
\end{equation}
one recovered node is chosen at random and becomes susceptible: R $\rightarrow$ S.

Finally, with probability
\begin{equation}
	P_{\text{IS}} = \frac{\lambda N_{\text{SI}}}{\mu\eta N_{\text{inf}} + \lambda N_{\text{SI}} + \alpha N_{\text{rec}}},
	\label{eq:GA_PIS}
\end{equation}
one infected node $i$ is selected with probability proportional to its degree. A nearest neighbor of $i$ is chosen at random and, if susceptible, becomes infected. If the neighbor is infected or recovered, no state alteration occurs, and the simulation proceeds to the next step. The time is incremented by 
\begin{equation}
	\delta t = \frac{-\ln u}{\mu\eta N_{\text{inf}} + \lambda N_{\text{SI}} + \alpha N_{\text{rec}}}
	\label{eq:GA_dt}
\end{equation}
where $u$ is a pseudo-random number uniformly distributed in the interval $(0,1)$. The SIS simulations use the same algorithm, replacing the transition I$_\eta \rightarrow$ R with I$_\eta \rightarrow$ S. 

To address some difficulties related to finite-size effects in absorbing state dynamics, we implemented the hub reactivation method~\cite{Guilherme2016}, where the node with the largest degree is reinfected whenever the number of infected individuals reaches zero.

\section{Saddle point method to compute Eq.~\eqref{eq:QKmed}}
\label{app:saddle}

To compute the asymptotic limit of Eq.~\eqref{eq:QKmed}, we define
\begin{equation}
Z_{\eta}(K)=
\alpha\int_{0}^{\infty}e^{-\alpha t_{2}}(1-p_1^2 p_2)^{K}dt_{2}. 
\label{eq:Zeta_eta}
\end{equation}
where  $p_1$, given by Eq.~\eqref{eq:P_1}, is a constant and $p_2$, according to Eqs.~\eqref{eq:Gamma} and \eqref{eq:Pinf}, is given by
\begin{equation}
\begin{aligned}
	p_2 & =P_\text{inf}(t_2)   
    	&   = 1-\frac{\widetilde{\gamma}(\eta,\eta\mu t_2)}{(\eta-1)!},
	\end{aligned}
\label{eq:p_2gammaincom}
\end{equation}
where $\widetilde{\gamma}$ is the incomplete Gamma function~\cite{zwillinger2007table} given by
\begin{equation}
	\widetilde{\gamma}(n,x) = \int_{0}^{x} y^{n-1} e^{-y}dy = (n-1)!\left[ 1-e^{-x}\sum_{m=0}^{n-1} \frac{x^{m}}{m!}\right]. 
	\label{eq:gammaincom}
\end{equation}
where the right-hand side is the series expansion of $\widetilde{\gamma}$~\cite{zwillinger2007table}.
Now, replacing $(1-p_1^2p_2)^K\simeq \exp(-Kp_1^2p_2)$, Eq.~\eqref{eq:p_2gammaincom}, and Eq.~\eqref{eq:gammaincom} in Eq.~\eqref{eq:Zeta_eta} we obtain
\begin{equation}
	Z_\eta(K) = \frac{\alpha}{\eta\mu}\int_{0}^{\infty} 
		\exp\left[-\frac{\alpha z}{\eta \mu} -Kp_1^2 e^{-z} \sum_{m=0}^{\eta-1} \frac{z^m}{m!}
	\right] dz,
	\label{eq:Z_eta_series}
\end{equation}
where we performed the change of variable $x=\eta\mu t_2$. Notice that Eq.~\eqref{eq:Z_eta_series} has the proper form to apply the saddle-point method~\cite{arfken2013mathematical}:
\begin{equation}
	Z_{\eta}(K)=\frac{\alpha}{\eta\mu}\int_{0}^{\infty}\exp\left(-f(z)\right)dz,
	\label{eq:Z_eta_saddle}
\end{equation}
where
\begin{equation}
	f(z)=\frac{\alpha z}{\eta\mu}+Kp_1^{2} e^{-z}\sum_{m=0}^{\eta-1}\frac{z^{m}}{m!}.
	\label{eq:f_de_z}
\end{equation} 
The extrema of Eq.~\eqref{eq:f_de_z}, $f'(z_*)=0$, are given by the solutions of transcendent equation
\begin{equation}
{z_*}^{\eta-1}e^{-z_*}=\frac{(\eta-1)!}{\eta}\frac{\alpha}{\mu Kp_1^{2}}.
	\label{eq:equ_z_star}
\end{equation}
For $\eta=1$, Eq.~\eqref{eq:equ_z_star} has a single solution which is a minimum:
\begin{equation}
	z_{*}=\ln\left(\frac{\mu Kp_1^{2}}{\alpha}\right).
	\label{eq:z_star_eta1}
\end{equation}
For $\eta\ge 2$, there are two solutions since $Kp_1^2\gg1$. Since 
\begin{equation}
	f''(z_{*})=\frac{\alpha}{\mu\eta}\left(1-\frac{\eta-1}{z_{*}} \right),
	\label{eq:fpp_de_z}
\end{equation}
the solution for small $z_*$ is a local maximum, not relevant for the saddle-point method. So, to the leading order in $K$, the minimum is then given by 
\begin{equation}
	z_{*} \simeq \ln\left(Kp_1^{2}\right)+(\eta-1)\ln\left[\ln\left(Kp_1^2\right) \right].
\end{equation}

The saddle-point methods consists of expanding $f(z)$ around its minimum $z_*$ up to second order and plug it into Eq.~\eqref{eq:Z_eta_saddle} to obtain
\begin{equation}
	Z_\eta(K) =  \frac{\alpha}{\eta\mu} e^{-f(z_*)}\sqrt{\frac{2\pi}{f''(z_*)}}.
	\label{eq:Z_eta_saddle2}
\end{equation}
Since $z_*\gg 1$, using Eq.~\eqref{eq:fpp_de_z}, we have $f''(z_*) \simeq\alpha/\mu\eta$ and, using Eqs.~\eqref{eq:f_de_z} and \eqref{eq:equ_z_star},  
\begin{equation}
	\begin{aligned}
f(z_*) & = \frac{\alpha z_*}{\eta\mu} +\frac{\alpha}{\mu\eta} \frac{(\eta-1)!}{z_*^{\eta-1}} \sum_{m=0}^{\eta-1} \frac{z_*^m}{m!}\\
       & = \frac{\alpha z_*}{\eta\mu}\left[ 1+ \frac{1}{z_*}+\mathcal{O}(z_*^{-1})\right] \simeq  \frac{\alpha z_*}{\eta\mu}.
\end{aligned}
\end{equation}
to the leading order. Therefore, introducing $f(z_*)$ and $f''(z_*)$ in Eq.~\eqref{eq:Z_eta_saddle2} leads to 
\begin{equation}
Z_{\eta}(K) \sim \exp\left(-\frac{\alpha z_*}{\mu\eta}\right) = 
[Kp_1^2]^{-\frac{\alpha}{\eta \mu}} \left[ \ln\left(Kp_1^2\right)\right]^{-\frac{\alpha(\eta-1)}{\eta\mu}}.
\end{equation}

\section{Homogeneous mean-field analysis for SI$_\eta$RS}
\label{app:RRN}

Since the transition between infectious stages are Poisson process, the set of mean-field equations for the SI$_\eta$RS  dynamics for $k$ contacts  is given by 
\begin{eqnarray}
	\frac{dI_1}{dt}&=&\lambda{k}(1-I-R)I-\mu\eta I_1, \nonumber\\
	\frac{dI_j}{dt}&=&\mu\eta I_{j-1} -\mu\eta I_j,~~j=2,\ldots,\eta \nonumber\\
	\frac{dR}{dt}&=&\mu\eta I_\eta -\alpha R,
	\label{eq:MF_SIRS_MULT}
\end{eqnarray}
where $I=\sum_{j=1}^{\eta} I_\eta$ and $S+I+R=1$ for a closed population. Performing a linear stability around the absorbing state fixed point $R=I_j=0$, we obtain the Jacobian matrix given by, 
\begin{equation}
\begin{small}	
\textbf{L}=\begin{pmatrix}
\lambda k-\mu\eta & \lambda k & \lambda k &\cdots & \lambda k & \lambda k & 0\\
\mu\eta & -\mu\eta& 0&\cdots & 0 & 0&0\\
0 &\mu\eta & -\mu\eta&\ddots & 0 & 0 & 0\\
\vdots & \ddots & &\ddots &-\mu\eta& 0 & 0\\		
0 & 0 & \cdots & 0 & \mu\eta  &-\mu\eta &0\\
0 & 0 & \cdots & 0 & 0& \mu\eta&-\alpha\\
\end{pmatrix}.
\end{small}
\end{equation}
The epidemic threshold corresponds to the value of $\lambda$ in which the largest eigenvalue of this matrix is null.
Note that, from a linear stability analysis, one can see that the Eq.\ref{eq:MF_SIRS_MULT} converges to the SI$_{\eta}$S model. Therefore, the results in Ref.~\cite{ROST2020} holds and the epidemic threshold is given by $\lambda_{\text{c}}=\mu/k$.

\begin{acknowledgments}
	
D.H.S. thanks the support given by \textit{Fundação de Amparo à Pesquisa do Estado de São Paulo} (FAPESP)-Brazil (Grants No. 2021/00369-0 and No. 2013/07375- 0).	
S.C.F. acknowledges the financial support by the \textit{Conselho Nacional de Desenvolvimento Científico e Tecnológico} (CNPq)-Brazil (Grant no. 310984/2023-8) and \textit{Fundação de Amparo à Pesquisa do Estado de Minas Gerais} (FAPEMIG)-Brazil (Grant No. APQ-01973-24). F.A.R. acknowledges CNPq (Grant No. 308162/2023-4) and FAPESP (Grant No. 19/23293-0) for the financial support given for this research. This study was financed in part by the \textit{Coordenação de Aperfeiçoamento de Pessoal de Nível Superior} (CAPES), Brazil, Finance Code 001.

\end{acknowledgments}


%

\end{document}